\documentclass[journal=mamobx,manuscript=article,layout=twocolumn] {achemso}
\usepackage{achemso,array,booktabs,lmodern}
\usepackage{amsmath}

\author{Athanassios Z. Panagiotopoulos}
\email{azp@princeton.edu}
\affiliation{Department of Chemical and Biological Engineering, Princeton University, Princeton, NJ 08544, U.S.A.}
\title{Solvent Selectivity controls Micro- versus Macrophase Separation in Multiblock Chains}

\begin{document}
\begin{figure}
    \includegraphics[width=3.25 in]{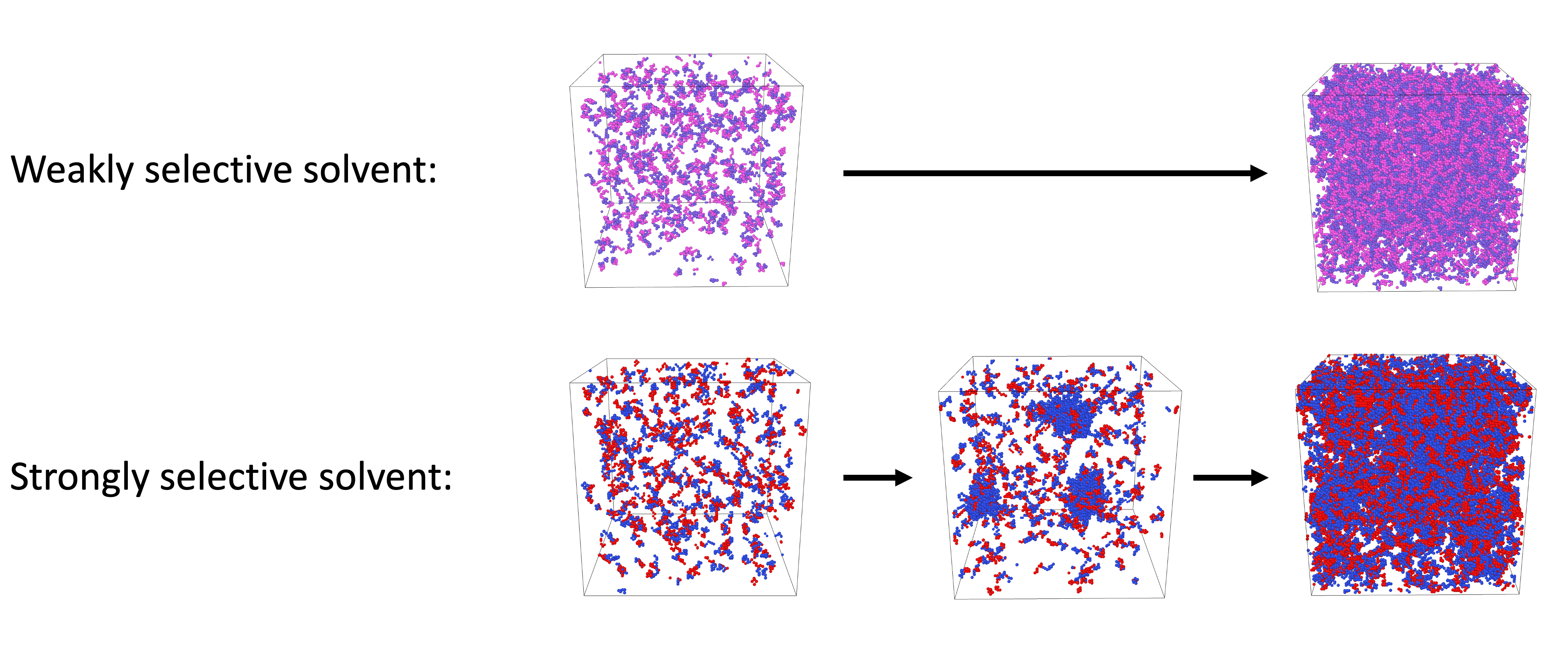}
    \caption*{Table of Contents Graphic}
 \label{fig:ToC}
 \end{figure}

\begin{abstract}
Monte Carlo simulations in the grand canonical ensemble were used to obtain critical parameters and conditions leading to microphase separation for block copolymers with solvophilic and solvophobic segments. Solvent selectivity was systematically varied to distinguish between systems that undergo direct macrophase separation and ones that initially microphase separate in the dilute phase. Finite-size scaling was used to obtain the critical parameters. Interestingly, corrections to scaling increase significantly for systems that form aggregates. The threshold value of solvent selectivity for aggregation was determined for symmetric diblock chains of varying length. The results suggest that long diblock copolymers form micelles in the dilute phase prior to macrophase separation, even in marginally selective solvents. The dependence of critical temperature on solvent selectivity was also obtained for triblock, multiblock, and alternating chains. For highly selective solvents, strong structuring in both dilute and dense phases makes it harder to reach equilibrium. 
\end{abstract}

\section{Introduction}

Formation of finite aggregates in solution, a special case of \emph{microphase} separation, is a continuous structural transition that is not associated with sharp discontinuities in thermodynamic functions. This is in contrast to \emph{macrophase} separation, which is a first-order thermodynamic transition that occurs when two or more distinct bulk phases have equal free energies at specific thermodynamic conditions. Of particular importance for technological applications are vapor-liquid and liquid-liquid transitions which are widely used for chemical separations.\cite{nob05} Such fluid-fluid transitions are associated with critical points, which correspond to stable limits of stability for the two coexisting phases and are dominated by fluctuations.\cite{Ma01} Critical points entail universal features common to all systems, that depend only on space dimensionality and the type of order parameter characterizing the transition. An example of such a universal feature is the scaling of the width of the coexistence curve with distance from the critical point -- the relevant ``critical exponent'' is $\beta = 0.326$ for the three-dimensional Ising universality class to which all common vapor-liquid and liquid-liquid critical points for fluids belong.\cite{Kim03} An interesting question that has received little prior attention in the literature is the connection between microphase separation on one hand, which has no ``universal'' features, and macrophase separation on the other, which follows precisely defined scaling relationships. The interplay between micro- and macrophase separation may also have a role in biological systems, in which intrinsically disordered proteins drive formation of biomolecular condensates.\cite{Bra09}

Block copolymers in solution are important systems with applications in areas such as drug delivery\cite{Jon03,Kat12} or synthesis of metal nanoparticles\cite{Rie03},  but are also interesting from the fundamental point of view of understanding molecular mechanisms and driving forces of self-assembly and microphase separation. A wide variety of chemical building blocks are available for their synthesis \cite{Had05}, which can be accomplished with fine control of block composition and molecular weight. The critical micellar concentration and associated thermodynamic quantities have been measured for many block copolymers.\cite{Ale94}. Solution structures formed by copolymer micelles have also been of significant interest.\cite{Lod23} Well-developed molecular field theories are available for phase transformations in block copolymer melts\cite{Coc06,Bat12} and also for their micellization behavior in solution.\cite{Ngu23}   

The existence of both macrophase and microphase separation for a single amphiphilic system has been observed experimentally for  copolymers with polyethylene oxide blocks in aqueous solutions\cite{Kos99,Cui19}, but that had generally been considered to originate in the special properties of water as a solvent. Some triblock polymers in non-aqueous media have been observed to undergo transitions from micellar to macrophase-separated states through chain bridging\cite{Veg01}. More recently, a detailed study of block-random copolymers in two non-aqueous solvents\cite{Tay24} found several examples of systems that form micelles in solution and also separate into macroscopic phases at higher concentrations or at lower temperatures.  

There have been many simulations of phase separation and aggregation of chains consisting of different monomer types.\cite{Tad20} With modern computational hardware, it is generally fairly straightforward to reach time scales needed for formation of aggregates in simulations of chains in poor solvents.\cite{Ano20} However, it remains a challenge to distinguish between finite-size aggregates that represent the true free-energy minimum of a system and formation of droplets from a bulk liquid phase. This has been a long-standing topic of research in the author's group, that was motivated by the realization that grand canonical Monte Carlo simulations of simple chain models can provide precise information on self-assembly into micelles \cite{Flo99,Kim01,Kim02}. This approach was subsequently extended\cite{Pan02} to show that some  architectures, specifically chains with short solvophilic segments, phase separate into normal liquids, while other sequences form micelles. A follow-up study for linear multiblock chains\cite{Gin08} found that increasing the overall chain length, or the length of segments of a single bead type, favors micellization over phase separation. More recently, a variety of chain sequences were determined to undergo aggregation or phase separation, depending on their blockiness and overall length.\cite{Ran21}  

Simulations of systems near critical points for macrophase separation transitions entail special considerations, given that the correlation length for density fluctuations becomes unbounded and thus cannot be contained in any finite simulation box. There is a well-defined framework for understanding the effects of truncation of correlations\cite{Bin84,Cha86} known as ``finite-size scaling.''  This requires calculations for a variety of simulation box sizes and extrapolation of the results to infinite size, using statistical-mechanical scaling relationships appropriate for the universality class of the transition of interest.  In general, these scaling relationships are only valid asymptotically, at the limit of large simulation boxes. ``Corrections to scaling'' are system-specific factors that control how quickly a certain system approaches the expected scaling behavior.\cite{Bin01} These corrections will turn out to play an important role in the present study.  

Until recently, simulated systems had  been observed to display only one type of transformation, either micellization or macrophase separation. The only prior study that clearly observed both transformations in a single system was Ref. \citenum{Pan23}, for two short triblock chain architectures with the solvophobic segments at their ends -- these can form bulk phases through chain bridging.  The main objective of the current work is to obtain insights into how one can observe a continuous evolution from macrophase separation to aggregation (microphase separation) and how the two types of transformation can be found at different thermodynamic conditions in a single system. The key to this evolution of phase separation and aggregation behavior turns out to be the solvent relative selectivity for the different bead types, which is a parameter previously neglected in simulation studies. 

\section{Model and Methods}

\subsection{Model}

The model used in this study, a modification of the lattice surfactant model of Larson \emph{et al.}, \cite{Lar85,Lar88} is summarized here for completeness. The model consists of linear chains of beads connected by bonds, each bead occupying a site on a simple cubic lattice. Bonds between adjacent beads can be in the [1 0 0], [1 1 0],  [1 1 1] and equivalent lattice directions, resulting in 26 possible connectivity vectors. There are two bead types, solvophobic and solvophilic, respectively labeled ``T'' and ``H'', for ``tail'' and ``head'' groups in the original surfactant-oriented version of the model. Non-bonded beads interact along the allowable connectivity directions with nearest neighbors up to a distance of $\sqrt{3}$ in lattice length units. There are no interactions with empty lattice sites, which can be considered to represent a monomeric solvent. The model has three independent energy parameters, $\epsilon_{\mathrm{TT}}$, $\epsilon_{\mathrm{HT}}$, and $\epsilon_{\mathrm{HH}}$, the units of which also set the temperature scale. In the original version of the model, normalized values of the interaction energies were set to $\epsilon_\mathrm{TT}=-1$ and $\epsilon_\mathrm{HT}=\epsilon_\mathrm{HH}=0$. Here, a generalized version with controllable solvent selectivity is used instead, while keeping the normalization condition $\epsilon_{\mathrm{TT}}+\epsilon_{\mathrm{HH}}=-1$:
\begin{equation} \label{eq:param} 
		\epsilon_\mathrm{TT} =  - \frac {1+c} 2 \ \ \  ;  \ \ \  \epsilon_\mathrm{HT} = \epsilon_\mathrm{HH}  =  - \frac {1-c} 2 
\end{equation}
Thus, the ``old'' version of the model uses $c=1$, which corresponds to a high degree of solvent selectivity for H over T beads. In principle, $c$ can be assigned any real value. For $c=0$, there is \emph{no} selectivity and the model reduces to the lattice homopolymer model studied in Ref. \citenum{Pan98}.  This is a useful limiting case to serve as a reference point. Values of  $c<0$ would make H more solvophobic than T, reversing the role of H and T beads, so are not of interest since they represent just a swapping of labels. Values of $c>1$ would make HT and HH interactions repulsive (positive). For these reasons, the present study is restricted to solvent selectivity parameter values of $0 \le c \le 1$, covering non-selective to strongly selective solvents. A prior study of critical parameters as a function of sequence for short chains\cite{Pan24} used $c=0.5$, corresponding to a weakly selective solvent.

As already implied in the Introduction, a key objective is to investigate the behavior of this family of models as a function of the parameter $c$, using accurate methods for determining free energies and critical points. The main questions are the dependence of critical parameters and fluid-phase separation boundaries on solvent selectivity strength, chain length, and overall chain architecture (e.g., diblock, triblock, multiblock, or alternating sequences). 

\subsection{Methods}

Computational methodologies are similar to those used previously.\cite{Pan23,Pan24} Monte Carlo simulations in the grand canonical ensemble in cubic boxes of varying edge length $L$ were performed to obtain critical points, phase coexistence curves, and critical micellar concentrations. The volume fraction when $N$ chains of length $r$ are present in the simulation box is defined as $\phi=Nr/L^3$. To facilitate insertion and removal of chains, an athermal Rosenbluth algorithm\cite{Ros55} was implemented as detailed in Ref. \citenum{Pan98}. Source codes, example input and output files, and tabulated values of the numerical results are available online, as explained in the  Data Availability Statement at the end of this article. Statistical uncertainties were calculated as the standard deviation of data from four runs at identical conditions, each with a different random number seed. 

For phase transitions with a critical point belonging to the three-dimensional Ising universality class, the expected order parameter distribution has been obtained as a convenient analytical expression by Tsypin and Bl\"ote:\cite{Tsy00} 
 \begin{equation} \label{eq:TB} 
		\mathcal{P}(x) =  \frac  {e^{-(0.7774 x^2-1)^2 (0.1228x^2+0.776)}} {2.3306} 
\end{equation}
Normalization constants have been included in Eq.~\ref{eq:TB} -- this was not the case in the original publication. The equation is symmetric about $x=0$ and it can be easily confirmed that $\mathcal{P}(x)$ has unit variance and integrates to unit area over $x$. For the Ising model which has up/down symmetry, the order parameter $x$ is strictly proportional to the total system magnetization. For molecular fluids that do not possess such symmetry, the mixed-field finite-size scaling method of Bruce and Wilding \cite{Bru92,Wil92} can be used to obtain a scalar order parameter as $X= N - sE$, where $E$ is the system energy, $N$ the number of chains in the simulation box, and $s$ is termed the ``field mixing'' parameter. From the unnormalized $X$ values, a normalized $x$ order parameter is computed by requiring that $\mathcal{P}(x)$ has zero mean and unit variance, and also integrates to unit area, same as for the universal order parameter distribution of Eq.~\ref{eq:TB}. 

Histogram reweighting with the Ferrenberg-Swendsen algorithm\cite{Fer89} was used to rescale data to nearby thermodynamic conditions, resulting in distributions $\mathcal{P}(N,E)$ as a function of the chemical potential of chains $\mu$ and temperature $T$. The critical temperature $T_\mathrm{c}$, critical chemical potential, $\mu_\mathrm{c}$, and field mixing parameter, $s$, were optimized by minimizing deviations between the observed distributions and the expected values from Eq.~\ref{eq:TB}.  

There are systematic effects of system size on the critical parameters.\cite{Ork99}  To address these effects, simulations were performed for at least three different system box sizes $L$ for each case studied and the critical parameters then extrapolated to infinite box size, $L \rightarrow \infty$, using the finite-size scaling relationships for the critical temperature and volume fraction:
\begin{equation} \label{eq:sc} 
\begin{split}
	T_\mathrm{c}(L) - T_\mathrm{c}(\infty) \propto L^{-(\theta+1)/\nu} \\ \phi_\mathrm{c}(L) - \phi_\mathrm{c}(\infty) \propto L^{-(1-\alpha)/\nu}
\end{split}	
\end{equation}
In these expressions, the values of the critical exponents appropriate to the three-dimensional Ising universality class are $\theta=0.54$, $\nu=0.629$, and $\alpha=0.11$.\cite{Ork99}  

Simulation runs consisting of $10^{8}$ to $5 \times 10^{9}$ insertion / removal attempts were used to determine the $\mathcal{P}(N,E)$ distributions near the expected critical conditions, with longer runs needed for larger systems in order to keep the number of Monte Carlo steps per particle roughly constant and also to ensure good sampling for aggregating systems. 
\begin{figure}
    \includegraphics[width=3.6 in]{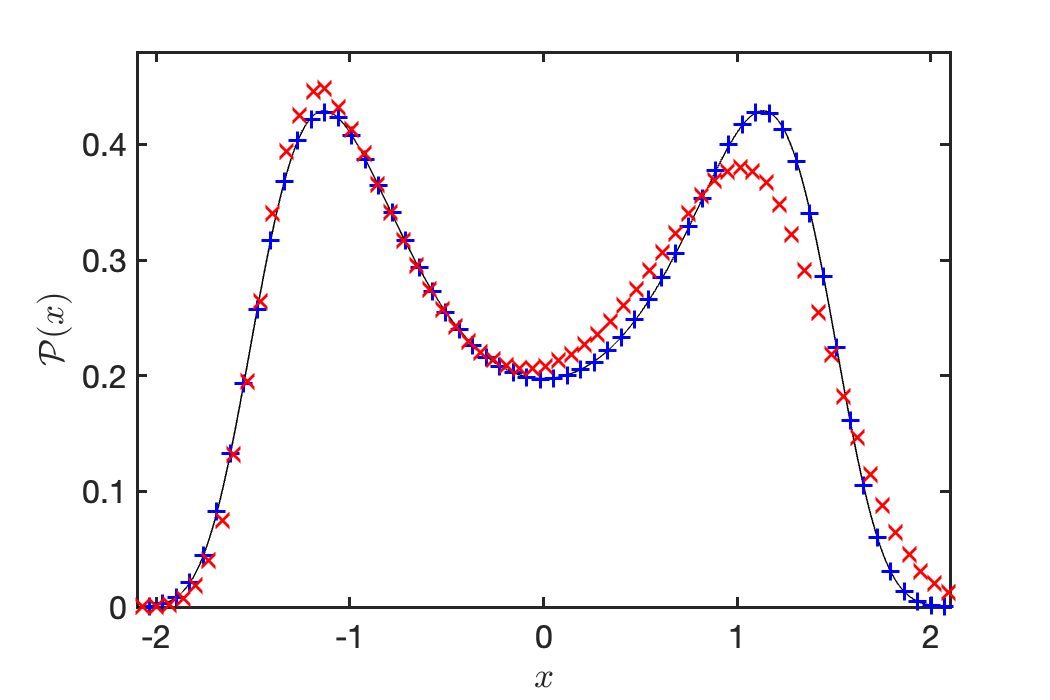}
    \caption{Order parameter distribution, $\mathcal{P}(x)$, versus $x$ at the estimated critical point for  H$_8$T$_8$, $L=34$; $c=0$ (blue $+$ symbols) and $c=0.5$ (red $\times$ symbols). The black line is from Eq. ~\ref{eq:TB}.} 
 \label{fig:P_x}
 \end{figure}

\section{Results and Discussion}
Typical simulation data for the order parameter distribution are shown in Fig.~\ref{fig:P_x} for two versions of the H$_8$T$_8$ system with different values of the solvent selectivity parameter $c$. There is nothing special about this particular chain architecture, as similar results were obtained for the other systems studied, but the use of relatively short chains facilitates efficient sampling. The analytical expression for the universal order parameter distribution of Eq.~\ref{eq:TB} is shown as a line for comparison. In both cases, simulation boxes had the same size ($L=34$) and the runs consisted of the same number of steps ($2 \times 10^9$). 

The discrepancies between observed data for the case $c=0.5$ (red $\times$ symbols) and the Ising universal distribution are not due to insufficient sampling, as the observed curve is fully converged with respect to run length at the scale of the graph, with statistical errors smaller than symbol size. Instead, these discrepancies are due to the formation of finite-size aggregates in both phases, which ``distort'' the order parameter distribution in a manner that mixed-field scaling cannot remove. As will be shown shortly,  $c=0.5$ is just beyond the threshold for formation of micellar aggregates prior to macroscopic phase separation in the H$_8$T$_8$ system. Differences between observed data at a finite $L$ and the universal distribution are termed corrections to scaling,\cite{Bin01} as described in the Introduction. They are expected to vanish at the limit of infinite box size. Corrections to scaling were indeed observed to become smaller as system size $L$ was increased. However, we cannot reach the limiting behavior for the order parameter distribution for this system, as it becomes impractical to sample much larger systems: the number of particles and therefore the number of required Monte Carlo steps for adequate sampling of the $\mathcal{P}(N,E)$ distributions scale as $L^3$. 
\begin{figure}   
\includegraphics[width=3.6 in]{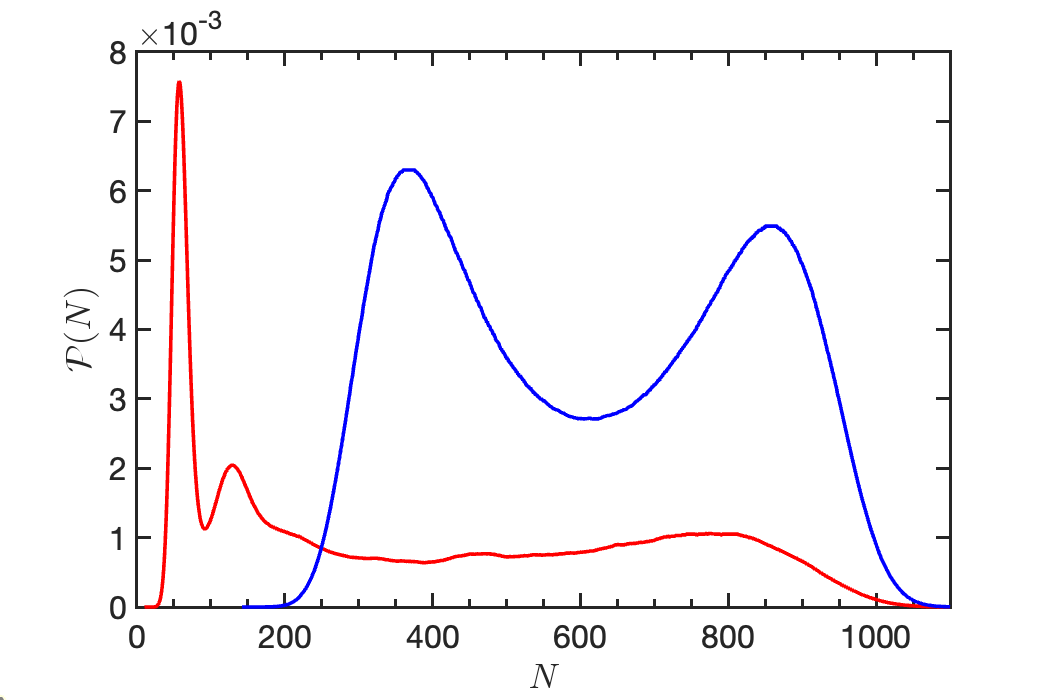}
    \caption{Probability distribution of the number of chains in the simulation box, $\mathcal{P}(N)$, versus $N$ for H$_8$T$_8$,  $L=34$; blue line corresponds to $c=0$ (multiplied by a factor of 3 for visual clarity) and red line to $c=0.5$. These distributions are at identical conditions to the transformed order parameter distributions (points) in Fig.~\ref{fig:P_x}.} 
 \label{fig:dens}
 \end{figure}

The appearance of stable micellar aggregates in systems such as H$_8$T$_8$, $c=0.5$, can be confirmed in multiple ways. One, already seen, is the discrepancy between the order parameter distribution and the Ising universal curve. More directly, the untransformed $\mathcal{P}(N)$ distribution develops two low-density peaks, as seen in Fig.~\ref{fig:dens} (red curve) for the same conditions as the transformed $\mathcal{P}(x)$ (red points) in Fig.~\ref{fig:P_x}. The first (highest) peak at  $N \approx 60$ corresponds to a dilute phase consisting primarily of monomers and oligomers, with no micellar aggregates. The second peak at  $N \approx 130$ corresponds to states with micellar aggregates in the dilute phase. The high-density peak at  $N \approx 800$ is very broad for this system and corresponds to a structured bulk liquid.  Transformation of the data using field mixing ($s=-0.017$) to obtain the red points in Fig.~\ref{fig:P_x} masks the double peak at low densities, but is unable to completely symmetrize the distribution. By contrast, data for $c=0$ (blue line) show a nearly-symmetric distribution of occupancies already at the $\mathcal{P}(N)$ level, and are fully symmetrized ($s=-0.0051$) to obtain the blue points in Fig.~\ref{fig:P_x} that are a close match to the universal order parameter distribution.  

\begin{figure}
\includegraphics[width=3.6 in]{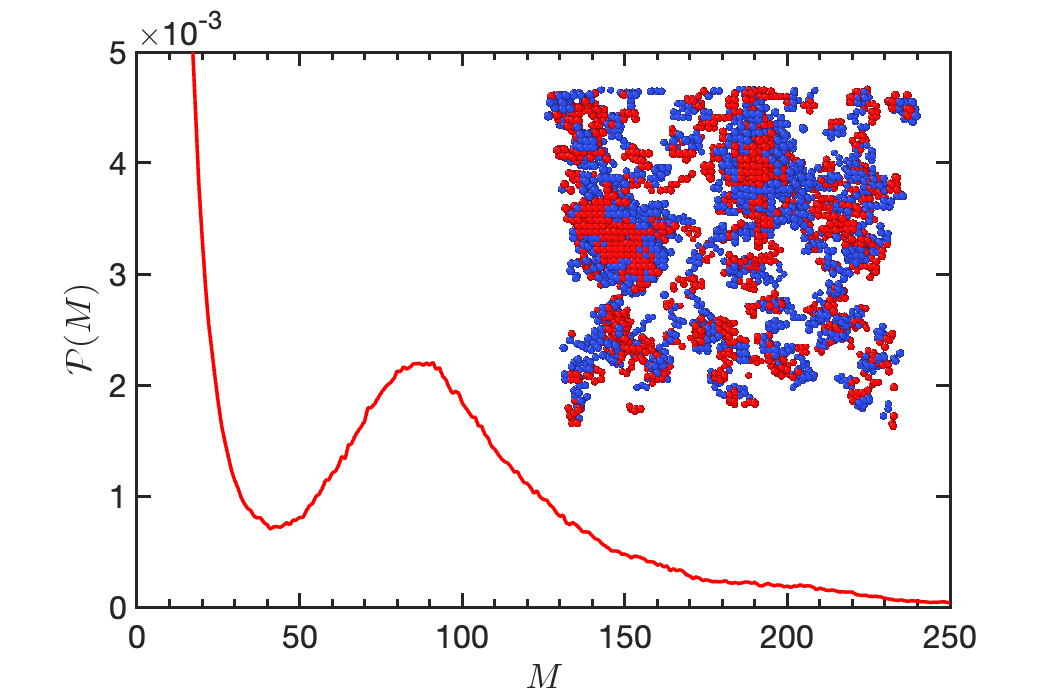}
 \caption{Cluster size distribution, $\mathcal{P}(M)$, versus cluster aggregation number $M$, for H$_8$T$_8$, $c=0.5$, $L=60$, at $T=4.855$ and $<\phi>= 0.025$.  The inset shows a representative simulation snapshot, with H beads in blue color and T beads in red.} 
 \label{fig:clu}
 \end{figure}
Fig.~\ref{fig:clu} displays a representative cluster size distribution for this system ($c=0.5$) at the extrapolated infinite-system critical temperature and a volume fraction a factor of 7 below the critical value. There is a clear micellar peak with favored aggregation number $M \approx 90$. Visual inspection of the corresponding configurations also shows clearly defined aggregates (inset of Fig.~\ref{fig:clu}), with somewhat irregular, non-spherical shapes because of the proximity to the critical point point. Based on prior studies of related systems\cite{Pan02}, micellar aggregates are expected to become more spherical and increase in size at lower temperatures.

The threshold value for appearance of finite-size aggregates in the low-density phase is designated from this point onwards as ``$c_\mathrm{m}$''. For any given chain architecture, it is obtained operationally by gradually increasing $c$ while testing for existence of a second low-density peak in the $\mathcal{P}(N)$ distribution at the estimated critical point for the largest size studied. 

\begin{figure}
    \includegraphics[width=3.7 in]{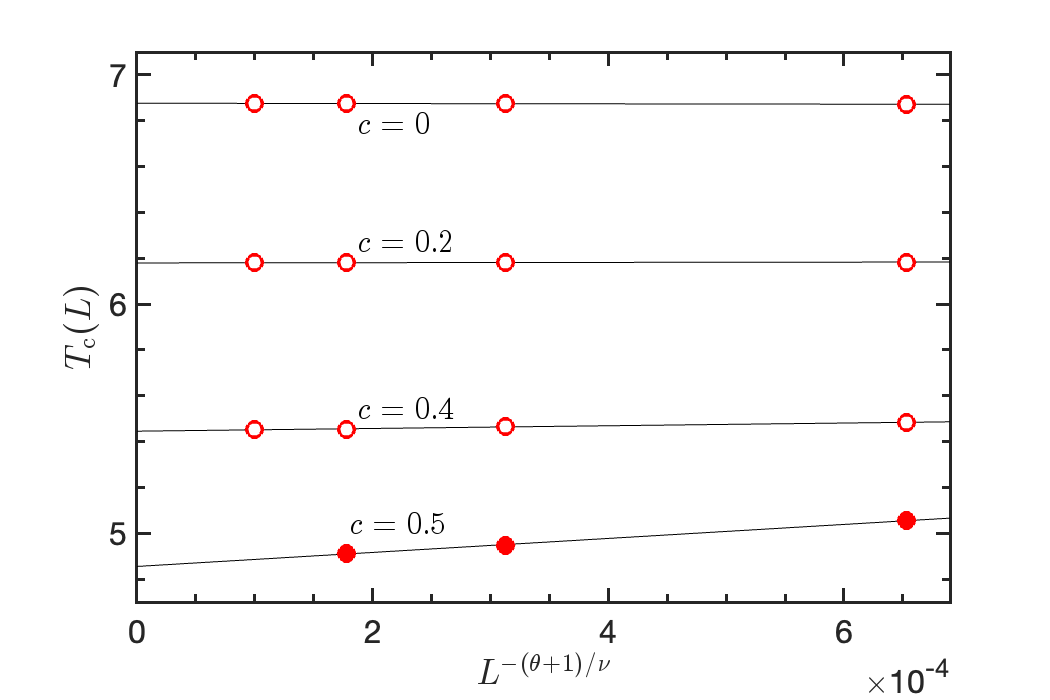}
    \caption{Scaling of the critical temperature $T_\mathrm{c}(L)$ with system size $L$ for H$_8$T$_8$ chains, following Eq.~\ref{eq:sc}. Lines are linear-least-squares fits to the points, with text labels giving the value of the solvent selectivity parameter $c$. Open symbols are systems with no aggregation in the dilute phase, while filled symbols are systems showing aggregation in the dilute phase. Error bars are much smaller than symbol size.} 
 \label{fig:H8L}
 \end{figure}
Once the critical parameters have been obtained for a given chain architecture, value of the solvent selectivity $c$, and system size $L$, it is desirable to extrapolate the results to infinite system size, $L \rightarrow \infty$. Fig.~\ref{fig:H8L} shows the critical temperatures of H$_8$T$_8$ chains with varying $c$, plotted against the relevant scaling function of $L$. Relative uncertainties for $T_\mathrm{c}(L)$ are better than $3 \times 10^{-4}$, much smaller than symbol size in this figure. Detailed numerical results for all systems are broadly similar and are available online as explained in the Data Availability Statement. As seen in the figure, the scaling relationship for the critical temperature is followed quite closely. For non-selective ($c=0$) or weakly selective solvents ($c=0.2$), the system-size dependence is not perceptible at the scale of the figure. However, for selective solvents ($c=0.4$) and especially for solvents with sufficient strength to induce microphase separation ($c=0.5$), there is a much stronger system size dependence, with the critical temperature dropping in the latter case by approximately 4\% from $L=20$ (smallest system studied, to the right of the figure) to the extrapolated value for $L \rightarrow \infty$ (intercept of the fitted lines with the ordinate axis). 

\begin{figure}   
    \includegraphics[width=3.6 in]{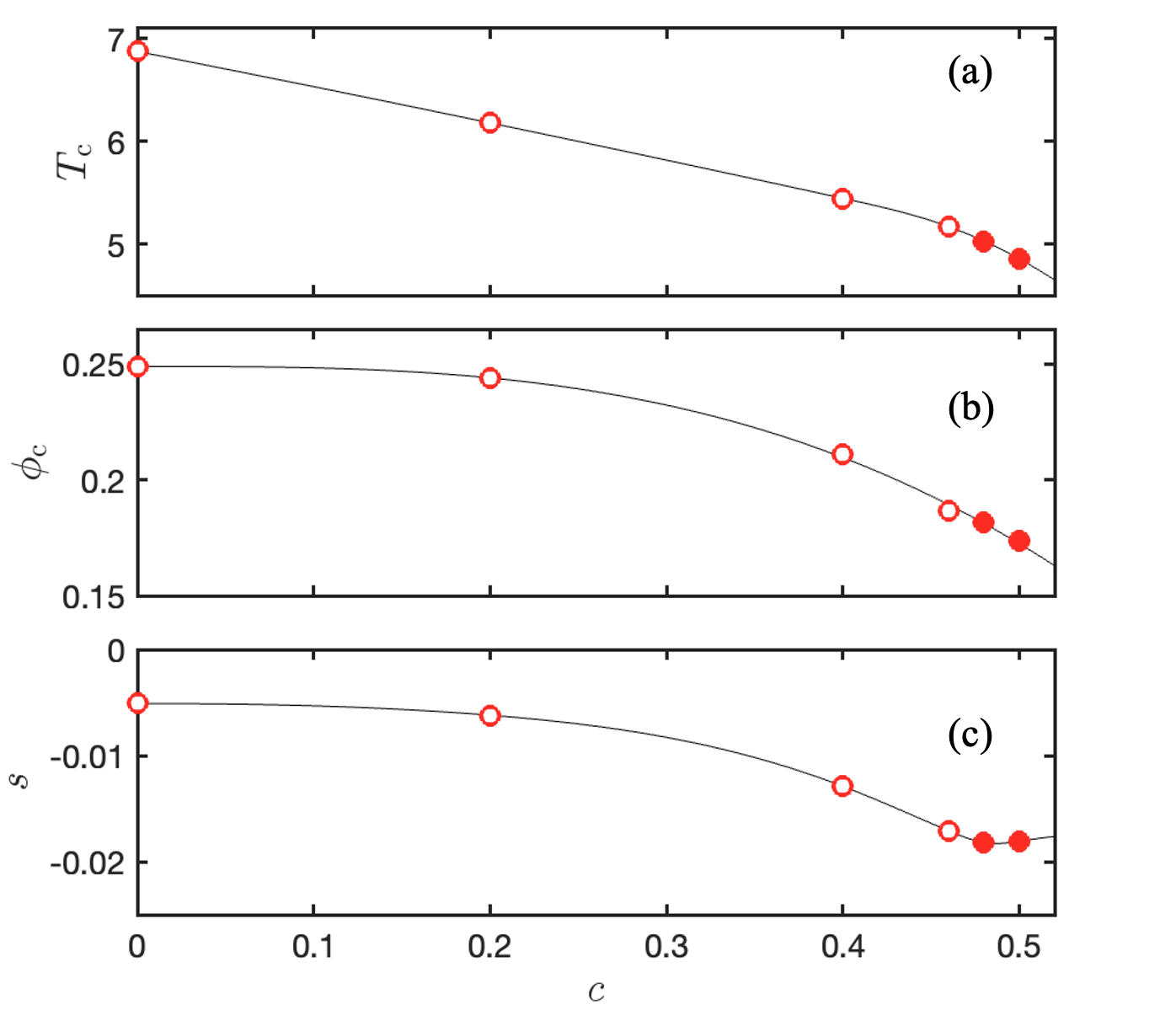}
    \caption{Dependence of (a) the critical temperature $T_\mathrm{c}$, (b) the critical volume fraction $\phi_\mathrm{c}$, and (c) the field mixing parameter $s$ on the solvent selectivity parameter $c$ for H$_8$T$_8$ chains. Open symbols are systems with no aggregation in the dilute phase, while filled symbols are systems showing aggregation in the dilute phase. Data have been extrapolated to infinite system size, $L \rightarrow \infty$. Lines are for guiding the eye. Error bars are smaller than symbol size. } 
 \label{fig:H8c}
 \end{figure}

The extrapolated infinite-system-size values of critical parameters for H$_8$T$_8$ chains are plotted in Fig.~\ref{fig:H8c} as a function of the solvent selectivity parameter $c$. The lines in this figure are only to guide the eye.  Points marked by filled symbols correspond to systems for which aggregates form in the dilute phase prior to macrophase separation, while open symbols correspond to non-aggregating systems. The critical temperature $T_\mathrm{c}$ is initially an almost-linear function of $c$, but the slope increases in the range of $c$ corresponding to aggregation. The critical volume fraction $\phi_\mathrm{c}$ and field mixing parameter $s$ behave in a similar way, being relatively unchanged initially when the solvent is weakly selective but changing faster as the aggregation limit is approached. For this system, the threshold solvent selectivity parameter for aggregation is $c_\mathrm{m} = 0.48 \pm 0.01$.  

\begin{figure}
 \includegraphics[width=3.6 in]{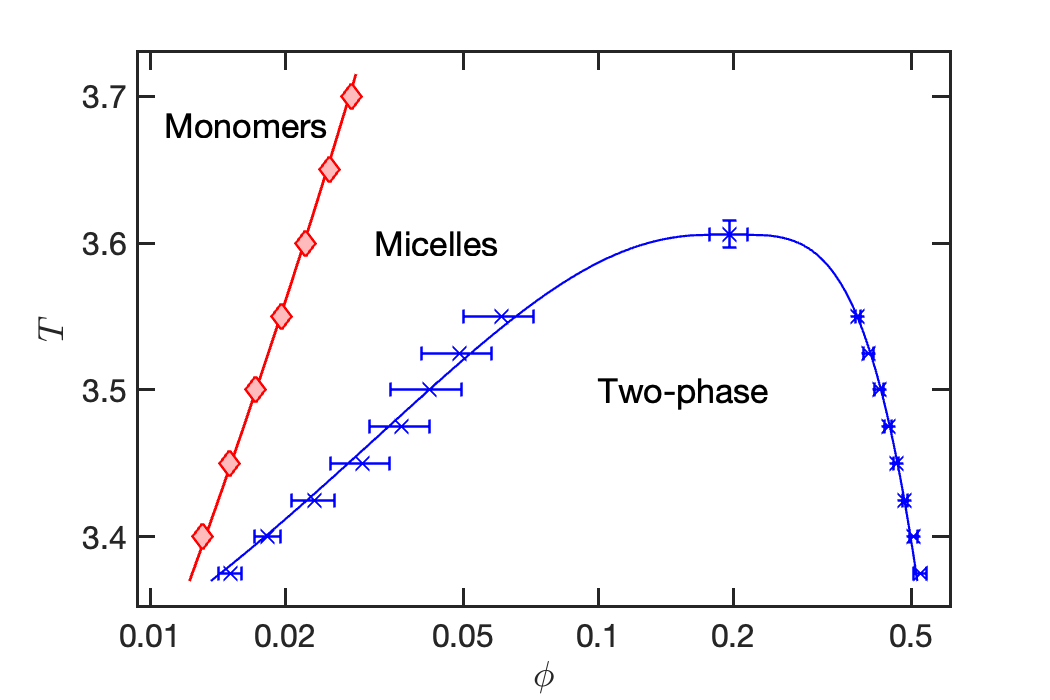}
    \caption{Temperature $T$ versus volume fraction $\phi$ for H$_4$T$_4$,  $c=0.66$, $L=32$. The binodal (equilibrium) points are shown as blue $\times$ symbols, with associated error bars. The lines through the binodal points were obtained as explained in the text. Critical micellar volume fractions are shown as red diamonds. The red line through them is a guide to the eye. Error bars for the critical micellar volume fractions are smaller than symbol size. } 
 \label{fig:cmc}
 \end{figure}

The phase and micellization behavior just beyond the threshold value of $c$ for aggregation is illustrated in Fig.~\ref{fig:cmc} for the H$_4$T$_4$ system, $c=0.66$, $L=32$. This system has a threshold value of solvent selectivity for micellization of $c_\mathrm{m}=0.64 \pm 0.01$. Shorter chains facilitate sampling of binodal curves below the critical point, but the qualitative features also apply to other systems forming micelles in the dilute phase. Note the logarithmic $\phi$ axis to allow for clear depiction of low volume fractions. 

To construct the phase diagram and micellization curves, simulations at subcritical conditions were necessary in addition to the runs near the critical point. The critical micellar concentration (cmc) line was obtained from the osmotic pressure versus concentration curves,\cite{Pan23} while the coexistence lines were obtained from the equal-area condition for $\mathcal{P}(N)$ on the two sides of the transition. The relatively large error bars for the coexistence volume fractions are due to the large system size used, necessary in order to observe aggregates in the dilute phase when performing the equal-area construction for the binodal. For smaller boxes, the binodal would be erroneously placed closer to the cmc line, because there is no opportunity for micelles to form. The cmc line has error bars smaller than symbol size because it only requires equilibration of the dilute gas phase with no micellar aggregates present. 

The scaling relationship for the coexistence curve width is $\phi_\mathrm{l} - \phi_\mathrm{g} = a( T _\mathrm{c} - T)^ \beta$, where $\phi_\mathrm{l}$ and $\phi_\mathrm{g}$ are the volume fractions of the two phases, $a$ is a fitted constant, and $\beta=0.326$ is the scaling exponent. This was used together with the approximate ``law of rectilinear diameters,'' $ \phi_\mathrm{l} + \phi_\mathrm{g} =  b( T _\mathrm{c} - T) + 2\phi_\mathrm{c}$, where $b$ is another fitted constant, to obtain analytical expressions for  $\phi_\mathrm{l}$ and $\phi_\mathrm{g}$ as functions of $T$, using the known values of  $T _\mathrm{c}$ and $\phi_\mathrm{c}$  for each system and separate optimizations of parameters $a$ and $b$ for $\phi_\mathrm{l}$ and $\phi_\mathrm{g}$. These are shown in the figure as the continuous line through the critical point. 

As shown in the figure, there is a region of temperatures between $T=3.4$ and $3.6$ for this system, for which micellization occurs at volume fractions between the cmc line and the  low-density phase boundary. Micellization is a  continuous structural (not first-order) transition and becomes less distinct at higher temperatures than the ones shown on the figure. There is also no clear distinction between an aggregated phase and the dense fluid at temperatures above the critical. This qualitative behavior is expected to occur for all systems that microphase separate in the dilute phase. However, for strongly aggregating systems, the macrophase separation critical point may be pushed to such low temperatures that equilibration of the relevant phases is not feasible. This is the reason that in most prior studies, this behavior was missed and systems were reported to be either micro- or macrophase separating. 

\begin{figure}    
    \includegraphics[width=3.6 in]{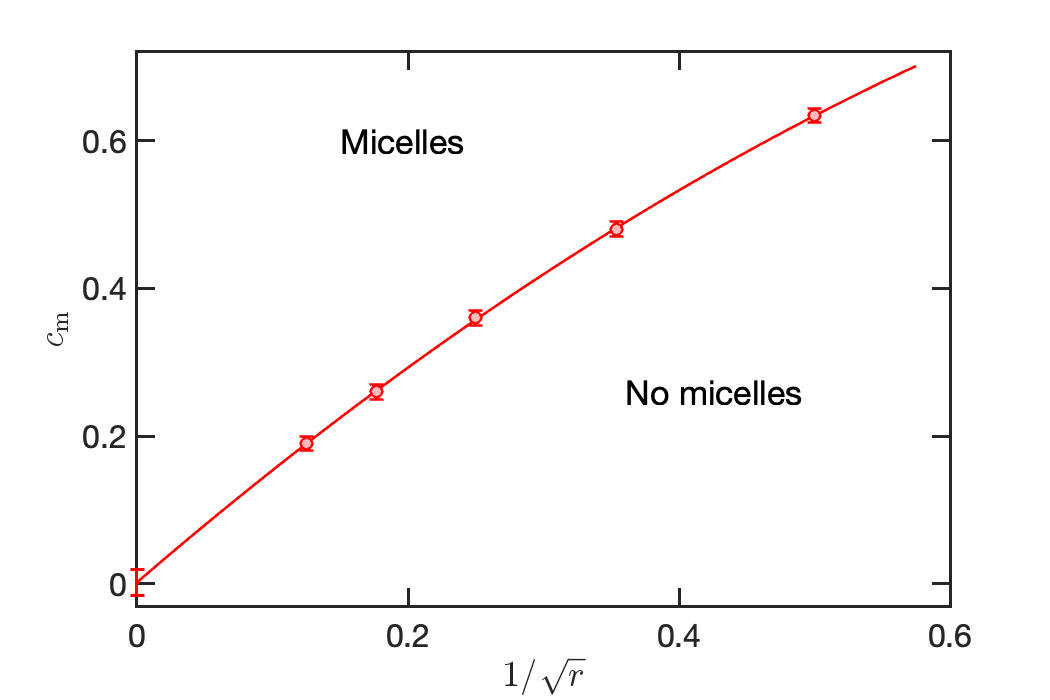}
    \caption{Solvent selectivity threshold for micellization, $c_\mathrm{m}$, as a function of the inverse square root of block size, $1/\sqrt{r}$, for diblocks of type H$_r$T$_r$. The values of $r$ for the points from left to right are 64, 32, 16, 8, and 4, respectively. The line is a quadratic polynomial fit to the data. Error bars are indicated on the data points as well as on the extrapolated intercept of the line with the ordinate axis.} 
 \label{fig:c_r}
 \end{figure}

It is clearly of interest to examine how the formation of aggregates depends on chain length. This key result from the present study is shown in Fig.~\ref{fig:c_r}. The threshold solvent selectivity parameter for formation of aggregates in the dilute phase, $c_\mathrm{m}$, is plotted in the figure against the inverse square root of block size, $1/\sqrt{r}$, for symmetric diblock chains of architecture H$_r$T$_r$.  Note that $1/\sqrt{r}$ is the dominant term for the scaling of the critical temperature with chain length according to the Schulz-Flory relationship\cite{Flo53} -- including the $1/2r$ second term in the Schulz-Flory relationship increases the non-linearity of the plot. The quantity $\sqrt r$ is proportional to the radius of gyration of the corresponding blocks in a $\Theta$ solvent. A good description of the data within their respective uncertainties is obtained using a quadratic polynomial fit. 
 
At any given block size (abscissa in Fig.~\ref{fig:c_r}), values of $c$ greater than the threshold $c_\mathrm{m}$ result in formation of micelles in the dilute phase prior to bulk phase separation. Conversely, values below the threshold $c_\mathrm{m}$ result in phase separation between a dilute phase lacking aggregates and a dense bulk liquid.  
From the quadratic fit shown, the intercept at $r \rightarrow \infty$ is $c_\mathrm{m} = 0.00 \pm 0.02$, which suggests that diblocks of infinite chain length aggregate to form micelles prior to phase separation even for weakly selective solvents. Recall that $c=0$ corresponds to a completely non-selective solvent. This result is complementary to the conclusion reached in Ref. \citenum{Ran21}, through an entirely different line of reasoning,  that only a small fraction of sequences aggregate prior to phase separation for increasingly long chains. Diblock and ``nearly-diblock'' sequences are the most aggregation-prone, but represent a minute fraction of all possible sequences for long chains with a fixed number of solvophilic and solvophobic beads. 

For systems at much higher value of $c$ relative to  $c_\mathrm{m}$, it becomes impractical to fully equilibrate the bulk liquid at temperatures near or below the critical point of the fluid-fluid transition, as also suggested earlier when discussing Fig.~\ref{fig:cmc}. The existence of highly structured, tightly bound aggregates renders insertion and removal moves less likely to be accepted. Even when they are accepted, these moves do not substantially change the overall morphology of the already-formed aggregates. Non-equilibrium structures are also likely present over experimental time scales for block copolymers in strongly selective solvents.\cite{Jon03,Mok12} 

Up to this point, the systems analyzed have all been symmetric diblock chains, consisting of two equal segments of solvophobic and solvophilic beads. It is of interest to investigate how these results carry over to other chain architectures. From prior studies\cite{Ran21}, it is clear that sequence strongly affects the tendency of a system to form finite-size aggregates.  In order to keep the focus on sequence effects (as opposed to chain length or composition effects), all chains studied in this part of the work consisted of 128 total beads and contained an equal number of solvophilic and solvophobic groups (50:50 overall composition).  
The architectures studied were the diblock H$_{64}$T$_{64}$ for which $c_\mathrm{m}$  has already been shown as the left-most data point in Fig.~\ref{fig:c_r}, the triblock H$_{32}$T$_{64}$H$_{32}$ that has solvophilic groups on the outside of the chain, its ``mirror image'' triblock, T$_{32}$H$_{64}$T$_{32}$ that has solvophobic groups on the outside of the chain, the octablock [H$_{16}$T$_{16}$]$_4$, and the alternating copolymer [HT]$_{64}$.  

\begin{figure}
    \includegraphics[width=3.7 in]{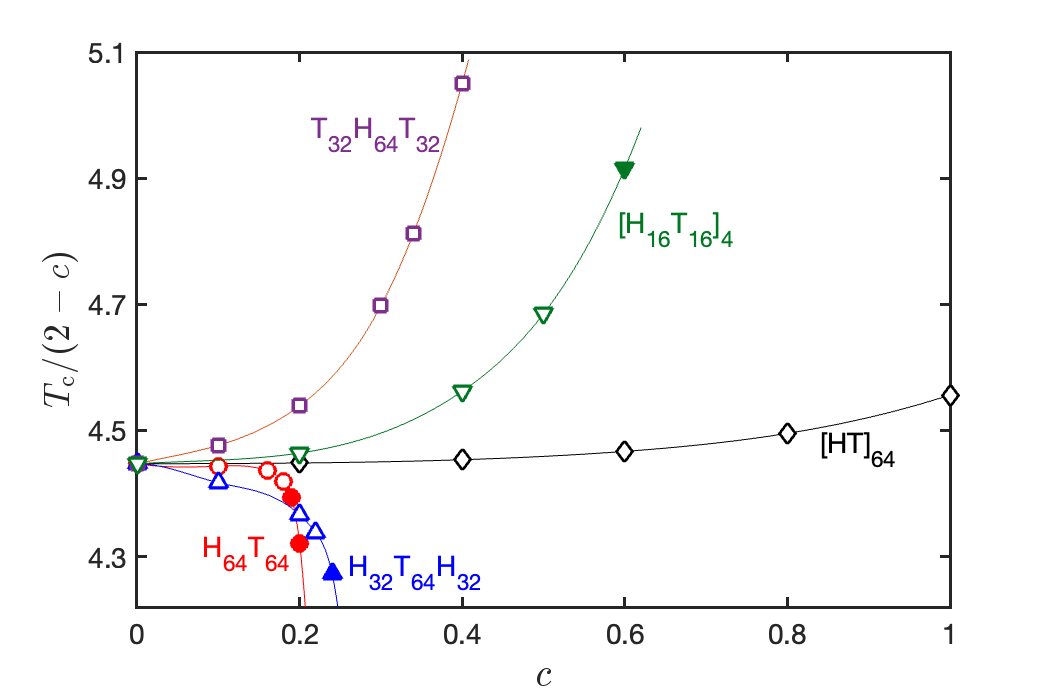}
    \caption{Normalized critical temperature, $T_\mathrm{c} / (2-c)$, versus $c$, for a range of chain architectures indicated by the text labels and corresponding colors. The lines through the points are to guide the eye. Open symbols are systems with no aggregation in the dilute phase, while filled symbols are systems showing aggregation in the dilute phase. Statistical uncertainties are smaller than symbol size} 
 \label{fig:c_128}
 \end{figure}

As already seen in Fig.~\ref{fig:H8c}(a), there is a significant decrease in the critical temperature for a given chain architecture as $c$ is increased. This is because HH and HT interactions in the model are identical and less attractive than TT interactions for all $c>0$. The critical temperature goes down with increasing $c$ because there are more HH and HT interactions than TT ones, both intramolecularly and intermolecularly. To highlight the effects of varying architecture, the temperature can be normalized by the linear factor $2-c$, which eliminates most of the variation of  $T_\mathrm{c}$ with $c$ for the alternating copolymer [HT]$_{64}$, as shown in Fig.~\ref{fig:c_128}. Unnormalized critical temperatures for this sequence vary by a factor of 2 in the $c$ range from 0 to 1, while the normalized ones differ by less than 3\%. 

The normalized critical temperature curves as a function of $c$ for all sequences are shown in Fig.~\ref{fig:c_128}. Contrary to the behavior of the alternating copolymer, there is a strong dependence of the normalized critical temperature on $c$ for the other sequences studied. All curves start at the same point for $c=0$, since at that limit H and T beads become identical and chain sequence becomes irrelevant.  

For any given value of $c$, the trends in the critical temperatures are consistent with what has been previously observed for short chains \cite{Pan24} based on the accessibility of blocks for interactions. Chains with solvophobic groups at their ends have higher critical temperatures than chains with uniform distribution of segments. Conversely, chains with solvophilic groups at the chain ends have lower critical temperatures. Diblock chains have critical temperatures close to the average value of uniformly distributed sequences, except when they approach the regions for which aggregation takes place in the dilute phase (filled points in Fig.~\ref{fig:c_128}). 

There is significant variation in the propensity of different sequences to form micellar aggregates.  Triblocks with external solvophobic chains (THT-type) require more selective solvents than triblocks with external solvophilic chains (HTH-type) to form micelles. This is because external solvophilic groups protect micelles from aggregation, while external solvophobic groups promote bulk phase separation. In addition, HTH-type chains can form traditional star-like micelles, whereas at low concentration the THT-type chains need to form flower-like micelles, more thermodynamically unfavorable because the H blocks need to form loops. For highly selective solvents, THT-type or multiblock systems form strongly structured bulk liquids in which domains of nearly pure solvophobic beads are connected via bridging chains, again leading to sampling difficulties. For such systems, formation of aggregates in the dilute phase prior to macrophase separation is observed at a relatively high value of the solvent selectivity parameter relative to diblock or HTH-type systems. For alternating copolymers, no aggregation is observed at any value of $c$, which is reasonable given that chain connectivity brings H segments into any incipient cluster of T beads. 

\section{Conclusions}
In the present work, a quantitative analysis was performed for the effects of solvent relative selectivity, chain length, and chain sequence on the microphase aggregation and macrophase separation behavior for block copolymers. Specifically, it was determined that in weakly selective solvents, macrophase separation between polymer-lean and polymer-rich phases takes place directly, in a single step. However, above a threshold value of the solvent relative selectivity, aggregation into micelles occurs in the dilute phase prior to phase separation. For solvent selectivities significantly higher than the threshold value, strong segregation in both dilute and condensed phases leads to slow equilibration. Symmetric diblock copolymers of increasing length require decreasing solvent selectivities to form micelles in the dilute phase prior to macrophase separation. The extrapolated infinite-chain-length limit suggests aggregation in the dilute phase prior to phase separation for long diblocks even in marginally selective solvents. Copolymer chains with solvophilic blocks at the ends show a decrease in their critical temperature with solvent selectivity, while chains with solvophilic blocks in the middle show the opposite effect. 

Macrophase separation between polymer-rich and polymer-lean phases is a first-order phase transition that belongs to the three-dimensional Ising universality class. The universal order parameter distribution at the critical point of the transition is closely followed for copolymers in weakly selective solvents that show little segregation in the dense phase. However, as solvent selectivity increases, corrections to scaling become larger, leading to significant deviations from the universal order parameter distribution at simulated system sizes for which it is possible to adequately sample these distributions. For more strongly selective solvents, formation of aggregates prevents the accurate determination of critical points using grand canonical Monte Carlo sampling. 

The connection between solvent selectivity and micro- or macrophase separation was most clearly elucidated experimentally in the very recent work of Taylor \emph{et al.}\cite{Tay24}, who obtained ``microphase separation into crew-cut micelles and macrophase separation into unimer and concentrated micelle phases [through] delicate tuning of the solvent/block compatibilities.''  While the specific numerical data obtained in the present study are for a generic lattice model of chains with solvophilic and solvophobic segments, it can be reasonably hypothesized that the qualitative findings are of general validity for synthetic or biological polymers with varying sequence, monomer chemical character, and solvent quality. For example, as seen in Ref. \citenum{Tay24}, the critical temperature increases for solvophilic segments in the interior of chains and decreases for solvophilic segments at the ends of chains, in agreement with the findings from the present study. In an ongoing project, we are using the model from the current work to obtain quantitative comparisons with experimental data for block-random copolymers, with results to be reported in a future publication. 

The findings from the present study have implications for the design of synthetic polymers with desired properties, specifically when aggregation in solution is a property of interest. For intrinsically disordered proteins in biological systems, liquid-liquid phase separation behavior is also an important consideration. The results suggest that there is a strong sensitivity of the overall phase separation and aggregation behavior to residue character, which effectively changes the solvent selectivity, as well as to protein sequence. In addition, results from the present work provide a conceptual framework explaining the recent discovery\cite{Kar22} of clusters in sub-saturated solutions of biomolecules that can undergo liquid-liquid phase separation. Solvent character, changed through addition of different anions, is found to play a role in controlling cluster formation and properties.\cite{Kar24}

\section*{Acknowledgements}

Financial support for this work was provided by the Princeton Center for Complex Materials (PCCM), a U.S. National Science Foundation Materials Research Science and Engineering Center (Award DMR-2011750). The author would like to thank Richard Register, Jerelle Joseph, Lauren Taylor, Ushnish Rana, and Amala Akkiraju for helpful discussions and comments on the manuscript. 

\section*{Data Availability Statement}
Computer codes used in this work, example input and output files, information on the runs performed, and numerical data for the critical points and phase coexistence curves are freely available for download from the Princeton Data Commons repository, at https://dx.doi.org/10.34770/yc39-ns70. 

\bibliography{library}

\end{document}